\documentclass[apj]{emulateapj}
\usepackage{apjfonts}
\usepackage{multirow}

\newcommand{\rl}{$R_{\rm BLR} - L$}
\newcommand{\msigma}{$M_{\rm BH}-\sigma_{\star}$}
\newcommand{\ml}{$M_{\rm BH}-L_{\rm bulge}$}
\newcommand{\mm}{$M_{\rm BH}-M_{\rm bulge}$}
\newcommand{\mbh}{$M_{\rm BH}$}
\newcommand{\sersic}{S\'{e}rsic}

\shorttitle{Low-Mass Black Hole in UGC\,06728}
\shortauthors{Bentz et al.}

\received{}
\accepted{}

\begin{document}

\title{A Low-Mass Black Hole in the Nearby 
  Seyfert Galaxy UGC\,06728}

\author{ Misty~C.~Bentz\altaffilmark{1},
Merida~Batiste\altaffilmark{1},
James~Seals\altaffilmark{1},
Karen~Garcia\altaffilmark{1},
Rachel~Kuzio~de~Naray\altaffilmark{1},
Wesley~Peters\altaffilmark{1},
Matthew~D.~Anderson\altaffilmark{1},
Jeremy~Jones\altaffilmark{1},
Kathryn~Lester\altaffilmark{1},
Camilo~Machuca\altaffilmark{1},
J.~Robert~Parks\altaffilmark{1},
Crystal~L.~Pope\altaffilmark{1},
Mitchell~Revalski\altaffilmark{1},
Caroline~A.~Roberts\altaffilmark{1},
Dicy~Saylor\altaffilmark{1},
R.~Andrew~Sevrinsky\altaffilmark{1},
and
Clay~Turner\altaffilmark{1}
}

\altaffiltext{1}{Department of Physics and Astronomy,
		 Georgia State University,
		 Atlanta, GA 30303, USA;
		 bentz@astro.gsu.edu}

\begin{abstract}

We present the results of a recent reverberation mapping campaign for
UGC\,06728, a nearby low-luminosity Seyfert 1 in a late-type galaxy.
Nightly monitoring in the spring of 2015 allowed us to determine an
H$\beta$ time delay of $\tau = 1.4 \pm 0.8$\,days.  Combined with the
width of the variable H$\beta$ line profile, we determine a black hole
mass of $M_{\rm BH} = (7.1 \pm 4.0) \times 10^5$\,M$_{\odot}$.  We
also constrain the bulge stellar velocity dispersion from
higher-resolution long slit spectroscopy along the galaxy minor axis
and find $\sigma_{\star} = 51.6 \pm 4.9$\,km\,s$^{-1}$.  The
measurements presented here are in good agreement with both the
\rl\ relationship and the \msigma\ relationship for AGNs.  Combined
with a previously published spin measurement, our mass determination
for UGC\,06728 makes it the lowest-mass black hole that has been fully
characterized, and thus an important object to help anchor the
low-mass end of black hole evolutionary models.

\end{abstract}


\section{Introduction}

Supermassive black holes are now believed to inhabit the nuclei of all
massive galaxies.  Furthermore, the active galactic nucleus, or AGN,
phase is generally understood to be a short-term event in the life of
a typical black hole, triggered either by a merger event or secular
processes in the host galaxy (cf.\ the review of \citealt{heckman14}
and references therein).  Tight scaling relationships between the
observed properties of black holes and their host galaxies point to a
symbiotic relationship between the two (e.g.,
\citealt{magorrian98,ferrarese00,gebhardt00,gultekin09,kormendy13,vandenbosch16}),
in which the growth of structure and the evolution of galaxies across
cosmic time is fundamentally linked to supermassive black holes.
Understanding this link requires an understanding of black hole
demographics, not just in the local universe, but also at higher
redshift where we can witness the growth of structure occurring.

Black holes, as opposed to galaxies, are incredibly simple objects
that can be fully characterized with only two fundamental
measurements: mass and spin.  In the Milky Way, years of astrometric
monitoring of stars in the central $\sim 0.01$\,parsec have led to an
extremely precise determination of the mass of our own supermassive
black hole \citep{ghez00,genzel00,ghez08}.  Unfortunately, all other
galaxies are too distant for this same technique to be employed, and
different techniques must be used to understand the masses of a
population of central black holes.  For galaxies out to $\sim
100$\,Mpc, spatially-resolved observations of the bulk motions of
stars or nuclear gas disks can be combined with dynamical modeling to
constrain the central black hole mass (cf.\ the reviews of
\citealt{ferrarese05,kormendy13}).  Reverberation mapping
\citep{blandford82,peterson93}, on the other hand, takes advantage of
AGN flux variability to constrain black hole masses through
time-resolved, rather than spatially-resolved, observations, thus
obviating any distance limitations.  Furthermore, the most widely-used
technique to constrain supermassive black hole spins requires high
X-ray luminosities that are only found in AGNs (e.g.,
\citealt{reynolds14} and references therein), so the study of active
black holes is an important key to unraveling the growth and evolution
of cosmic structure.

Unfortunately, bright AGNs are relatively rare in the local universe,
leading to a disconnect in our current understanding of nearby black
holes compared to those observed at larger look-back times.  In
particular, we are lacking direct comparisons of black hole mass
constraints through multiple independent techniques in the same
galaxies.  There are a handful of published comparisons of
reverberation masses and gas dynamical masses (e.g.,
\citealt{hicks08}), including the low-mass Seyfert NGC\,4395
(\citealt{peterson05,denbrok15}).  The agreement is generally
quite good, although the number of galaxies studied is small.  Stellar
dynamics, on the other hand, is a good check against reverberation
masses because it relies on modeling a non-collisional system, unlike
gas dynamics where the AGN may be expected to inject energy on
resolvable spatial scales.  However, only two such comparisons
currently exist for black hole masses from reverberation mapping and
stellar dynamical modeling: NGC\,4151 \citep{bentz06b,onken14} and
NGC\,3227 \citep{denney09c,davies06}.  While the techniques give
roughly consistent masses for these two examples, there are caveats
and limitations to both reverberation mapping and dynamical modeling,
and a larger comparison sample is needed to fully assess the
consistency of the local and the cosmological black hole mass scales.
We have therefore undertaken a program to identify and monitor local
AGNs where it might be possible to obtain both a reverberation and a
stellar dynamical mass constraint.  Both techniques are time- and
resource-intensive, and there are very few broad-lined AGNs within
$z\lesssim 0.01$, where the spatial resolution provided by $8-10$-m
class telescopes would be likely to resolve the black hole's
gravitational influence on the nuclear stellar dynamics, but we hope
to increase the sample of mass comparisons by a factor of a few.  We
currently have stellar dynamical modeling underway for two other local
AGNs, and we describe here the reverberation results for an additional
local AGN in our sample, UGC\,06728.

\section{Observations}
UGC\,06728 is a low-luminosity Seyfert 1 located at
$\alpha=$11:45:16.0, $\delta=+$79:40:53, $z=0.00652$ in a late-type
galaxy that is highly inclined to our line of sight.  It was
monitored nightly over the course of two months in the spring of 2015.
Optical spectroscopy and photometry were obtained at Apache Point
Observatory in New Mexico, with additional supporting photometry
obtained at Hard Labor Creek Observatory in Georgia.  We describe the
details below.

\subsection{Spectroscopy}

\begin{figure}
\epsscale{1.1}
\plotone{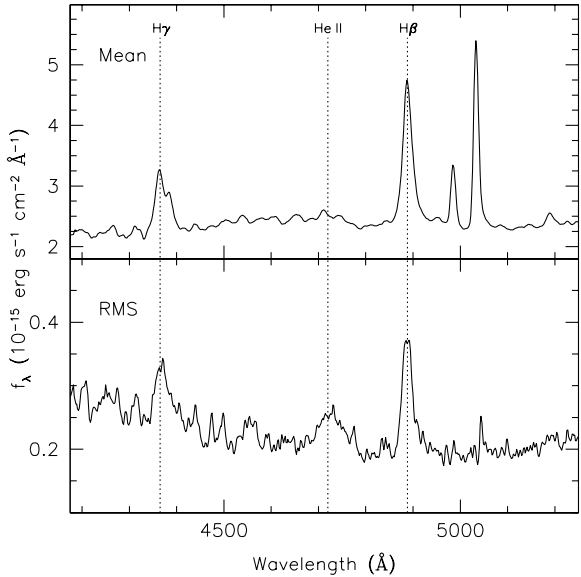}
\caption{Mean ({\it top}) and root mean square ({\it bottom}) of all
  the blue-side spectra obtained from APO during the monitoring campaign.}
\label{fig:mean.rms}
\end{figure}

Spectrophotometric monitoring of UGC\,06728 was carried out at Apache
Point Observatory (APO) with the 3.5-m telescope from 2015 April 15
$-$ May 30 (UT dates here and throughout).  Our monitoring program was
scheduled for the first hour of almost every night during this time
period, coincident with evening twilight.  We employed the Dual
Imaging Spectrograph (DIS), which uses a dichroic to split the
incoming beam into a red arm and a blue arm, with the low-resolution
(B400/R300) gratings centered at 4398\,\AA\ and 7493\,\AA.  The B400
and R300 gratings, when used together, cover the entire optical
bandpass between the atmospheric cutoff and 1\,$\mu$m, with a nominal
dispersion of 1.8\,\AA/pix and 2.3\,\AA/pix respectively.  Spectra
were obtained through a 5\arcsec\ slit rotated to a position angle of
0\degr (oriented north-south) and centered on the AGN.  On each visit,
a single spectrum with an exposure time of 600\,s was acquired at a
typical airmass of 1.5.  Observations of the spectrophotometric
standard star Feige\,34 were also acquired with each visit.

All spectra were reduced with IRAF\footnote{IRAF is distributed by the
  National Optical Astronomy Observatory, which is operated by the
  Association of Universities for Research in Astronomy (AURA) under
  cooperative agreement with the National Science Foundation.}
following standard procedures.  An extraction width of 12\,pixels was
adopted, corresponding to an angular width of 5\arcsec\ and
4.8\arcsec\ for the blue and red cameras, respectively.

The desire to minimize sampling gaps and maximize temporal coverage
means that ground-based reverberation campaigns must rely on
spectroscopy obtained under nonphotometric conditions.  While a
spectrophotometric standard star can help correct the overall shape of
the spectrum for atmospheric effects, as well as those from the
telescope and instrument optics, an additional technique is required
to achieve absolute flux calibrations of all the
spectra. Fortuitously, the narrow emission lines do not vary on short
timescales of weeks to months, so they can serve as convenient
``internal'' flux calibration sources.  We utilize the
\citet{vangroningen92} spectral scaling method, which accounts for
small differences in wavelength calibration, flux calibration, and
resolution (from variations in the seeing).  The method compares each
spectrum to a reference spectrum built from the best spectra
(identified by the user) and minimizes the differences within a
specified wavelength range.  The method has been shown to result in
relative spectrophotometry that is accurate to $\sim 2$\%
\citep{peterson98a}.  We restricted the scaling algorithm to focus on
the spectral region containing the [\ion{O}{3}]\,
$\lambda\lambda$\,4959,5007 doublet.  Additionally, we adopted an
overall flux scale based on the integrated [\ion{O}{3}]\,$\lambda5007$
flux measured from the nights with the best observing conditions of
$f_{\lambda5007} = 41.6\times10^{-15}$\,ergs\,s$^{-1}$\,cm$^{-2}$.
The red-side spectra showed only H$\alpha$ emission smoothly blended
with [\ion{N}{2}]\,$\lambda\lambda$\,6548,6583. Emission from
[\ion{S}{2}]\,$\lambda\lambda$\,6716,6730 and
[\ion{O}{1}]\,$\lambda\lambda$\,6300,6363 is extremely weak and
difficult to detect above the continuum.  With no suitable narrow
lines available, we were unable to accurately intercalibrate the
red-side spectra and we do not consider them further.

Figure~\ref{fig:mean.rms} displays the final mean and root mean square
(rms) of all the calibrated spectra acquired throughout the campaign.
The rms spectrum displays the variable spectral components, of which
H$\beta$, \ion{He}{2} $\lambda$\,4686, and H$\gamma$ are apparent, as
is the AGN continuum.  

\begin{figure}
\epsscale{0.7}
\plotone{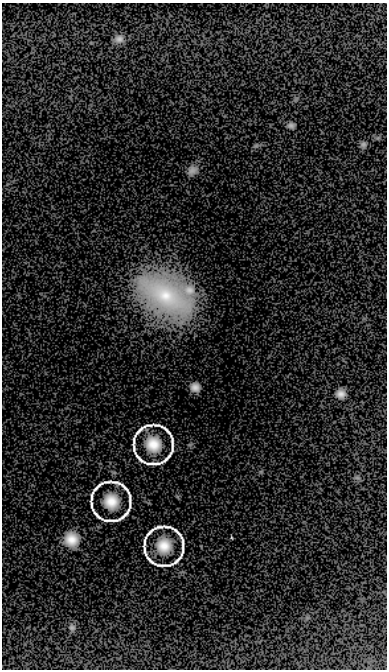}
\caption{Example $r-$band image acquired with the imaging mode of the
  DIS spectrograph at APO.  The field stars used to set the magnitude
  zeropoint are marked with circles.  The scale of the image is
  3\farcm9$\times$6\farcm7 and is oriented with north up and east to
  the right.}
\label{fig:apo.r}
\end{figure}

\subsection{Photometry}

Broad-band $g$ and $r$ images were obtained at APO with the imaging
mode of the DIS spectrograph each night directly after acquiring the
spectra.  The dual-arm nature of the spectrograph allowed both images
to be obtained simultaneously.  The typical exposure time was 30\,s,
and a single image in each filter was obtained per visit.  Images were
reduced in IRAF following standard procedures.  The DIS imaging mode
provides a relatively small field of view ($\sim
4$\arcmin$\times$7\arcmin), but there were a handful of convenient
bright stars in all of the images (see Figure~\ref{fig:apo.r}).  We
carried out aperture photometry employing circular apertures with
radii of 3\farcs78 in $g$ and 3\farcs6 in $r$, and sky annuli of
6\farcs3$-$7\farcs56 and 6\farcs0$-$7\farcs2 respectively.  Calibrated
$g-$ and $r-$band magnitudes for three field stars were adopted from
APASS (the AAVSO Photometric All Sky Survey; \citealt{henden14}) and
set the photometric zeropoints.

Photometric monitoring was also carried out with the 24-inch Miller
Telescope at Hard Labor Creek Observatory (HLCO), owned and operated
by Georgia State University in Hard Labor Creek State Park near
Rutledge, GA.  $V-$band images were acquired with an Apogee
$2048\times2048$ detector, spanning a field of view of $26\farcm3
\times 26\farcm3$ with a pixel scale of $0\farcs77$.  On a typical
night, three exposures were obtained at an airmass of $\sim 1.5$, each
with an exposure time of 300\,s.

The wide field of view of the HLCO images included a large number of
field stars, allowing us to derive a $V-$band light curve for
UGC\,06728 by employing image subtraction techniques.  We first
registered all the images to a common alignment with the {\tt Sexterp}
package \citep{siverd12}.  We then carried out the image subtraction
analysis with the {\tt ISIS} package \citep{alard98,alard00}.  {\tt
  ISIS} builds a reference frame from the best images (specified by
the user) and then uses a spatially-varying kernel to convolve the
reference frame to match each individual image in the dataset.
Subtraction of the two results in a residual image in which all
constant components have disappeared and only variable flux remains.
In the case of UGC\,06728, the host-galaxy and the average AGN
brightness are subtracted from all the residual images, leaving behind
only the brightness of the AGN relative to its mean level.  Aperture
photometry is then employed to measure this variable flux, which may
be positive or negative, at the location of the target of interest in
each residual image, providing a $V-$band residual light curve.

\begin{figure}
\epsscale{1.1}
\plotone{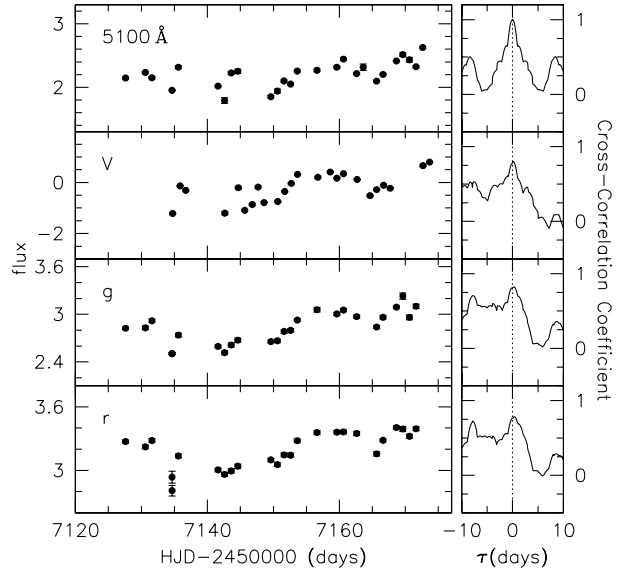}
\caption{Spectroscopic continuum and photometric light curves ({\it
    left panels}) and the cross-correlation of each light curve
  relative to the spectroscopic continuum light curve ({\it right
    panels}).  No apparent time delays are detected, except perhaps in
  the $r$ band, and the light curve features are quite similar.}
\label{fig:lc.cont}
\end{figure}

\begin{deluxetable*}{lcccc}[h!]
\tablecolumns{5}
\tablewidth{0pt}
\tablecaption{Photometric Light Curves}
\tablehead{
\colhead{HJD} &
\colhead{$g$} &
\colhead{$r$} &
\colhead{HJD} &
\colhead{$V$} 
\\
\colhead{(days)} &
\colhead{(AB mag)} &
\colhead{(AB mag)} &
\colhead{(days)} &
\colhead{(resid. cts/10000)}
}
\startdata
7127.6059 & $15.546 \pm	0.006$ & $14.823 \pm 0.005$ & 7134.7311 & $ 1.216 \pm  0.024$ \\
7130.6023 & $15.544 \pm	0.008$ & $14.840 \pm 0.006$ & 7135.8689 & $ 0.133 \pm  0.025$ \\
7131.6043 & $15.510 \pm	0.008$ & $14.820 \pm 0.006$ & 7136.7193 & $ 0.309 \pm  0.024$ \\
7134.6484 & $15.676 \pm	0.006$ & $14.941 \pm 0.021$ & 7142.6382 & $ 1.202 \pm  0.060$ \\
7134.6493 & $15.677 \pm	0.005$ & $14.989 \pm 0.020$ & 7144.6886 & $ 0.205 \pm  0.042$ \\
7135.6058 & $15.580 \pm	0.009$ & $14.869 \pm 0.006$ & 7145.6625 & $ 1.093 \pm  0.034$ \\
7141.6148 & $15.638 \pm	0.007$ & $14.916 \pm 0.006$ & 7146.8040 & $ 0.865 \pm  0.038$ \\
7142.6108 & $15.671 \pm	0.010$ & $14.931 \pm 0.007$ & 7147.6872 & $ 0.180 \pm  0.045$ \\
7143.6100 & $15.630 \pm	0.011$ & $14.919 \pm 0.007$ & 7148.5865 & $ 0.789 \pm  0.021$ \\
7144.6099 & $15.605 \pm	0.011$ & $14.903 \pm 0.007$ & 7150.6653 & $ 0.746 \pm  0.024$ \\
7149.6224 & $15.612 \pm	0.006$ & $14.882 \pm 0.005$ & 7151.7214 & $ 0.351 \pm  0.026$ \\
7150.6151 & $15.608 \pm	0.009$ & $14.898 \pm 0.006$ & 7152.7171 & $ 0.036 \pm  0.020$ \\
7151.6161 & $15.561 \pm	0.008$ & $14.866 \pm 0.006$ & 7153.6520 & $-0.319 \pm  0.022$ \\
7152.6159 & $15.556 \pm	0.008$ & $14.867 \pm 0.006$ & 7156.7405 & $-0.204 \pm  0.042$ \\
7153.6223 & $15.507 \pm	0.006$ & $14.820 \pm 0.005$ & 7158.6132 & $-0.410 \pm  0.025$ \\
7156.6189 & $15.460 \pm	0.010$ & $14.795 \pm 0.006$ & 7159.6255 & $-0.170 \pm  0.026$ \\
7159.6214 & $15.479 \pm	0.007$ & $14.794 \pm 0.005$ & 7160.6317 & $-0.346 \pm  0.009$ \\
7160.6215 & $15.461 \pm	0.007$ & $14.793 \pm 0.005$ & 7162.6928 & $-0.118 \pm  0.021$ \\
7162.6218 & $15.491 \pm	0.007$ & $14.798 \pm 0.006$ & 7164.6410 & $ 0.514 \pm  0.025$ \\
7165.6511 & $15.540 \pm	0.008$ & $14.862 \pm 0.007$ & 7165.6504 & $ 0.280 \pm  0.025$ \\
7166.6239 & $15.494 \pm	0.008$ & $14.819 \pm 0.005$ & 7166.7008 & $ 0.109 \pm  0.024$ \\
7168.6555 & $15.448 \pm	0.005$ & $14.780 \pm 0.005$ & 7167.7025 & $ 0.226 \pm  0.021$ \\
7169.6224 & $15.400 \pm	0.014$ & $14.785 \pm 0.007$ & 7172.6709 & $-0.661 \pm  0.037$ \\
7170.6218 & $15.495 \pm	0.011$ & $14.807 \pm 0.006$ & 7173.6752 & $-0.801 \pm  0.030$ \\
7171.6249 & $15.444 \pm	0.009$ & $14.784 \pm 0.006$ & & 
\label{tab:lc.phot}
\enddata 
\end{deluxetable*}

\section{Light Curve Analysis}

Light curves for the broad emission lines H$\beta$,
\ion{He}{2}\,$\lambda 4686$, and H$\gamma$ were derived directly from
the scaled spectra.  We fit a local, linear continuum below each
emission line and then integrated the flux above this continuum to
determine the total emission-line flux.  This includes the
contribution from the narrow component of each emission line, which is
simply a constant flux offset.  We also determined a continuum light
curve from the spectra at $5100\times(1+z)$\,\AA, which has the merit
of being completely uncontaminated by emission lines.  The strong
continuum and emission-line variability over the course of the
campaign allows us to determine these light curves directly from the
spectra without carrying out any spectral modeling or decomposition,
which has the potential to introduce artificial features into light
curves.

In Figure~\ref{fig:lc.cont}, we show the spectroscopic continuum light
curve relative to the $V-$band residual light curve and the $g$ and
$r$ photometric light curves (tabulated in Table~\ref{tab:lc.phot}).
The $V-$band residual light curve does not contain significant
emission from any broad emission lines, so we combined it with the
continuum light curve determined from our spectra to improve the time
sampling, especially in the first half of the campaign.  We selected
pairs of points from the two light curves that were contemporaneous
within $0.5$\,days and fit for the best multiplicative and additive
factors to bring the $V-$band residual fluxes into agreement with the
measured continuum flux densities. These best-fit factors account for
the differences in host-galaxy background light, average AGN flux
level, and bandpass.  The $V-$band light curve was scaled according to
the best-fit parameters and merged with the continuum light curve.  We
then examined the $g-$band light curve from the APO photometry and
found that there was no significant time delay relative to the merged
continuum+$V$ light curve, so we merged it as well by again finding
the multiplicative and additive scale factors necessary to bring it
into agreement with contemporaneous points in the continuum+$V$ light
curve.  Our final merged continuum light curve was binned to 0.5\,day
sampling to improve the accuracy.  The overall shape of the $r-$band
light curve agrees with the other photometric light curves and the
continuum light curve, but the variability level is somewhat damped by
additional host-galaxy flux and there is possibly a slight delay in
the light curve, so we did not merge the $r-$band with the other light
curves.  A detectable delay in $r$ is not unexpected, given that the
filter bandpass is centered on H$\alpha$.  While $g$ is centered on
H$\beta$, the overall contribution of H$\beta$ to the total filter
bandpass is much smaller than for H$\alpha$ and $r$.  In particular,
H$\beta$ contributes only 2\% of the $g-$band flux, with the variable
component of H$\beta$ accounting for only 10\% of the total H$\beta$
contribution, or 0.2\% of the total $g-$band flux.  On the other hand,
H$\alpha$ contributes 15\% of the total $r-$band flux.

\begin{figure}
\epsscale{1.1}
\plotone{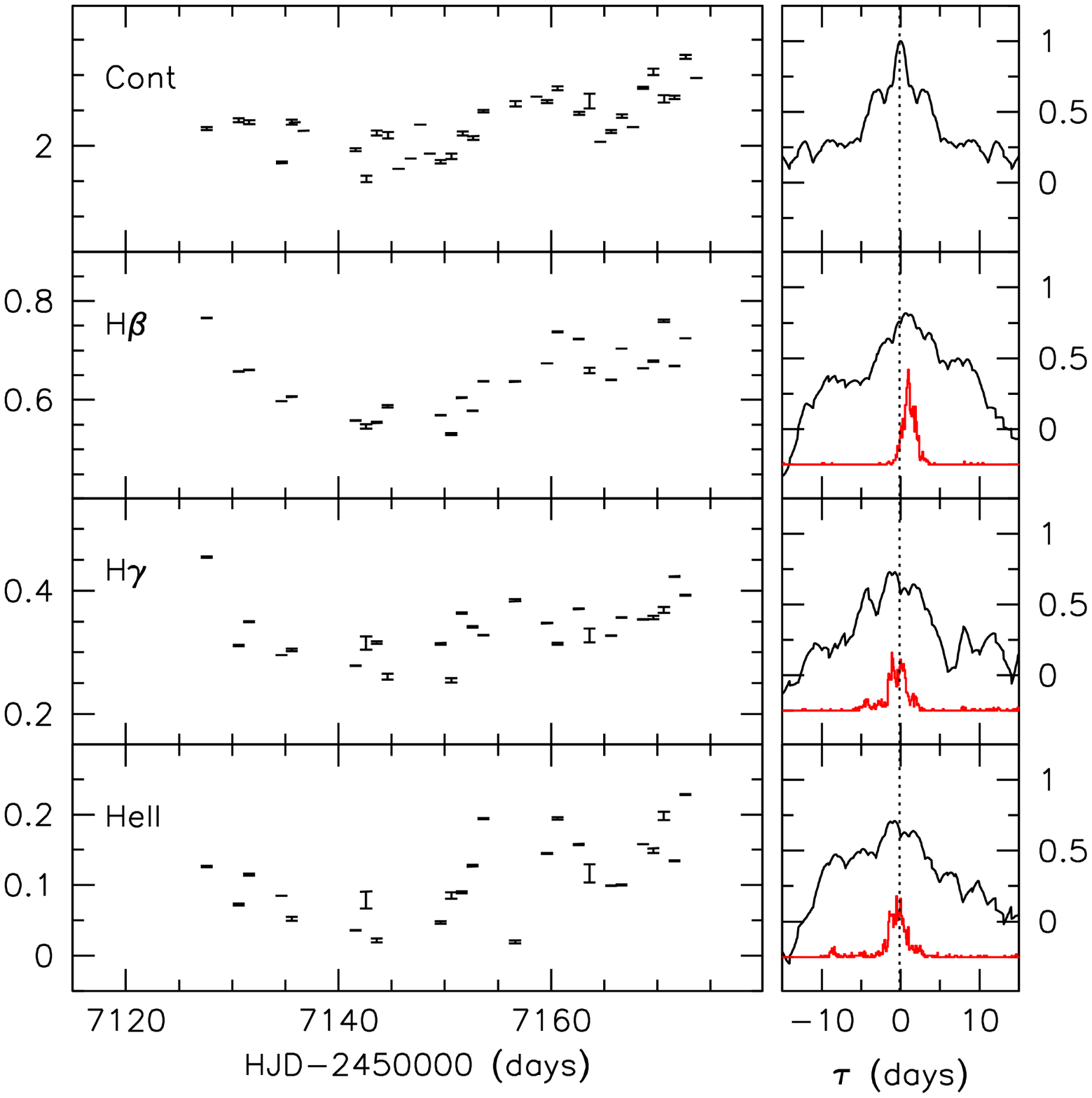}
\caption{Merged continuum light curve and emission-line light curves
  ({\it left panels}).  The right panels display the cross-correlation
  of each light curve relative to the continuum, and the red
  histograms (arbitrarily scaled) display the cross-correlation
  centroid distributions.}
\label{fig:lc}
\end{figure}

\begin{deluxetable*}{lcccc}
\tablecolumns{5}
\tablewidth{0pt}
\tabletypesize{\scriptsize}
\tablecaption{Spectroscopic Light Curves}
\tablehead{
\colhead{HJD} &
\colhead{$5100\times(1+z)$\,\AA} &
\colhead{H$\beta$} &
\colhead{H$\gamma$} &
\colhead{\ion{He}{2}} 
\\
\colhead{(days)} &
\colhead{($10^{-15}$\,erg\,s$^{-1}$\,cm$^{-2}$\,\AA$^{-1}$)} &
\colhead{($10^{-15}$\,erg\,s$^{-1}$\,cm$^{-2}$)} &
\colhead{($10^{-15}$\,erg\,s$^{-1}$\,cm$^{-2}$)} &
\colhead{($10^{-15}$\,erg\,s$^{-1}$\,cm$^{-2}$)}
}
\startdata
7127.60131629 &	$2.146	\pm 0.013$ & $76.605 \pm 0.039$ & $45.467 \pm 0.091$ & $12.614 \pm 0.092$ \\
7130.59772753 &	$2.235	\pm 0.018$ & $65.745 \pm 0.064$ & $31.085 \pm 0.126$ & $7.226  \pm 0.158$ \\
7131.59975985 &	$2.151	\pm 0.016$ & $66.059 \pm 0.053$ & $34.986 \pm 0.098$ & $11.465 \pm 0.127$ \\
7134.64244313 &	$1.953	\pm 0.007$ & $59.745 \pm 0.009$ & $29.567 \pm 0.011$ & $8.501  \pm 0.014$ \\
7135.60113042 &	$2.317	\pm 0.023$ & $60.694 \pm 0.101$ & $30.367 \pm 0.206$ & $5.223  \pm 0.263$ \\
7141.6102368  &	$2.019	\pm 0.011$ & $55.835 \pm 0.025$ & $27.859 \pm 0.045$ & $3.583  \pm 0.056$ \\
7142.60623451 &	$1.792	\pm 0.045$ & $54.629 \pm 0.464$ & $31.498 \pm 1.061$ & $7.873  \pm 1.207$ \\
7143.6053809  &	$2.227	\pm 0.023$ & $55.439 \pm 0.101$ & $31.572 \pm 0.221$ & $2.160  \pm 0.262$ \\
7144.60529841 &	$2.253	\pm 0.033$ & $58.688 \pm 0.236$ & $26.036 \pm 0.502$ & \nodata \\ 
7149.61781551 &	$1.852	\pm 0.021$ & $56.921 \pm 0.089$ & $31.359 \pm 0.145$ & $4.714  \pm 0.176$ \\
7150.61051989 &	$1.941	\pm 0.031$ & $53.080 \pm 0.185$ & $25.462 \pm 0.354$ & $8.538  \pm 0.455$ \\
7151.61149262 &	$2.101	\pm 0.016$ & $60.457 \pm 0.054$ & $36.374 \pm 0.109$ & $8.983  \pm 0.131$ \\
7152.61133458 &	$2.051	\pm 0.016$ & $57.825 \pm 0.050$ & $34.132 \pm 0.107$ & $12.753 \pm 0.123$ \\
7153.61769047 &	$2.256	\pm 0.011$ & $63.765 \pm 0.025$ & $32.758 \pm 0.042$ & $19.423 \pm 0.055$ \\
7156.61430815 &	$2.268	\pm 0.021$ & $63.760 \pm 0.086$ & $38.456 \pm 0.175$ & $1.949  \pm 0.212$ \\
7159.61682615 &	$2.317	\pm 0.013$ & $67.404 \pm 0.033$ & $34.757 \pm 0.066$ & $14.462 \pm 0.080$ \\
7160.61690956 &	$2.445	\pm 0.019$ & $73.758 \pm 0.067$ & $31.379 \pm 0.128$ & $19.432 \pm 0.164$ \\
7162.61714264 &	$2.216	\pm 0.013$ & $72.343 \pm 0.031$ & $37.067 \pm 0.058$ & $15.755 \pm 0.076$ \\
7163.62052311 &	$2.317	\pm 0.051$ & $65.958 \pm 0.505$ & $32.747 \pm 1.106$ & $11.655 \pm 1.309$ \\
7165.63733328 &	$2.098	\pm 0.010$ & $64.085 \pm 0.019$ & $32.687 \pm 0.026$ & $9.924  \pm 0.032$ \\
7166.61939046 &	$2.203	\pm 0.013$ & $70.381 \pm 0.032$ & $35.662 \pm 0.059$ & $10.006 \pm 0.076$ \\
7168.65095798 &	$2.415	\pm 0.008$ & $66.380 \pm 0.012$ & $35.326 \pm 0.015$ & $15.805 \pm 0.019$ \\
7169.61788092 &	$2.515	\pm 0.026$ & $67.836 \pm 0.125$ & $35.631 \pm 0.273$ & $14.877 \pm 0.332$ \\
7170.61722538 &	$2.434	\pm 0.037$ & $76.008 \pm 0.228$ & $36.909 \pm 0.465$ & $19.810 \pm 0.575$ \\
7171.62035657 &	$2.325	\pm 0.014$ & $66.832 \pm 0.036$ & $42.273 \pm 0.072$ & $13.417 \pm 0.088$ \\
7172.62633782 &	$2.627	\pm 0.014$ & $72.460 \pm 0.035$ & $39.271 \pm 0.062$ & $22.804 \pm 0.080$ 
\label{tab:lc.spec}
\enddata 
\end{deluxetable*}

\begin{deluxetable*}{lccccccc}
\tablecolumns{8}
\tablewidth{0pt}
\tablecaption{Light-Curve Statistics}
\tablehead{
\colhead{Time Series} &
\colhead{$N$} &
\colhead{$\langle T \rangle$} &
\colhead{$T_{\rm median}$} &
\colhead{$\langle F \rangle$\tablenotemark{a}} &
\colhead{$\langle \sigma_F/F \rangle$} &
\colhead{$F_{\rm var}$} &
\colhead{$R_{\rm max}$}\\
\colhead{} &
\colhead{} &
\colhead{(days)} &
\colhead{(days)} &
\colhead{} &
\colhead{} &
\colhead{} &
\colhead{}
}
\startdata
5100\,\AA  & 26 & $1.8 \pm 1.4$ & 1.0 & $2.21 \pm 0.20$ & 0.009 & 0.090 & $1.466 \pm 0.038$ \\
$V$        & 24 & $1.7 \pm 1.3$ & 1.1 & $-0.22 \pm 0.56$ & 0.050 & -2.56 & $-0.659 \pm 0.028$ \\
$g$        & 25 & $1.8 \pm 1.4$ & 1.0 & $2.83 \pm 0.21$ & 0.008 & 0.072 & $1.290 \pm 0.017$ \\
$r$        & 25 & $1.8 \pm 1.4$ & 1.0 & $3.19 \pm 0.17$ & 0.007 & 0.052 & $1.213 \pm 0.023$ \\
H$\beta$   & 26 & $1.8 \pm 1.4$ & 1.0 & $64.3 \pm 6.7$  & 0.002 & 0.105 & $1.443 \pm 0.005$ \\
H$\gamma$  & 26 & $1.8 \pm 1.4$ & 1.0 & $33.9 \pm 4.6$  & 0.007 & 0.135 & $1.786 \pm 0.025$ \\
\ion{He}{2}& 25 & $1.9 \pm 1.5$ & 1.0 & $11.3 \pm 5.7$  & 0.033 & 0.500 & $11.7 \pm 1.3$ 
\label{tab:lcstats}
\enddata 

\tablenotetext{a}{5100\,\AA, $g-$band, and $r-$band flux densities are
  in units of $10^{-15}$\,ergs\,s$^{-1}$\,cm$^{-2}$\,\AA$^{-1}$.
  $V-$band residual flux is in units of 10000\,counts.  Emission
  line fluxes are in units of $10^{-15}$\,ergs\,s$^{-1}$\,cm$^{-2}$.}

\end{deluxetable*}

Figure~\ref{fig:lc} displays the final merged and binned continuum
light curve and the broad emission-line light curves (tabulated in
Table~\ref{tab:lc.spec}.  The variability statistics for each of the
light curves are tabulated in Table~\ref{tab:lcstats}.  Column (1)
lists the spectral feature and column (2) gives the number of
measurements in the light curve.  Columns (3) and (4) list the average
and median time separation between measurements, respectively.  Column
(5) gives the mean flux and standard deviation of the light curve, and
column (6) lists the mean fractional error (based on the comparison of
observations that are closely spaced in time).  Column (7) lists the
excess variance, computed as:
\begin{equation}
F_{\rm var} = \frac{\sqrt{\sigma^2 - \delta^2}}{\langle F \rangle}
\end{equation}
\noindent where $\sigma^2$ is the variance of the fluxes, $\delta^2$
is their mean-square uncertainty, and $\langle F \rangle$ is the mean
flux. And column (8) is the ratio of the maximum to the minimum flux
in the light curve, $R_{\rm max}$.

We employed the interpolated cross-correlation function (ICCF)
methodology \citep{gaskell86,gaskell87} with the modifications of
\citet{white94} to search for time delays of the emission lines
relative to the continuum.  The ICCF method calculates the
cross-correlation function (CCF) twice, by interpolating first one
light curve and then the other, and averages the two results together
to determine the final CCF.  The CCF can be characterized by its
maximum value ($r_{\rm max}$), the time delay at which the maximum
occurs ($\tau_{\rm peak}$) and the centroid ($\tau_{\rm cent}$) of the
points around the peak above some value (typically $0.8 r_{\rm max}$).
CCFs for each light curve relative to the continuum are displayed in
Figure~\ref{fig:lc} ({\it right panels}).  For the continuum light
curve, this is the autocorrelation function.  

To quantify the uncertainties on the time delay measurements,
$\tau_{\rm cent}$ and $\tau_{\rm peak}$, we employ the Monte Carlo
``flux randomization/random subset sampling'' method of
\citet{peterson98b,peterson04}. This method is able to account for the
measurement uncertainties as well as the effect of including or
excluding any particular data point.  The ``random subset sampling''
is implemented such that, from the $N$ available data points within a
light curve, $N$ points are selected without regard to whether a point
has been previously chosen.  For a point that is sampled $1 \leq n
\leq N$ times, the uncertainty on that point is scaled by a factor of
$n^{1/2}$.  The typical number of points that is not selected in any
specific realization is $\sim 1/e$.  The ``flux randomization''
component takes the newly sampled light curve and modifies the flux
values by a Gaussian deviation of the flux uncertainty.  These
modified light curves are then cross-correlated with the ICCF method
described above, and the whole process is repeated many times
($N=1000$). From the large set of realizations, we build distributions
of $\tau_{\rm cent}$ and $\tau_{\rm peak}$.  The median of each
distribution is taken to be the measurement value, and the
uncertainties are set such that they mark the upper 15.87\% and lower
15.87\% of the realizations (corresponding to $\pm 1\sigma$ for a
Gaussian distribution).  The red histograms in the Figure~\ref{fig:lc}
depict the cross-correlation centroid distribution for each emission
line.

To further check that combining the various photometric and
spectroscopic light curves has not affected our measured time delays,
we also determined the time delay of H$\beta$ relative to each of the
individual continuum, $V-$band, and $g-$band light curves.  Each of
these light curves is slightly undersampled relative to the combined
continuum light curve, but the CCFs and recovered H$\beta$ time delays
agree within the measurement uncertainties.

\begin{figure}
\epsscale{1.1}
\plotone{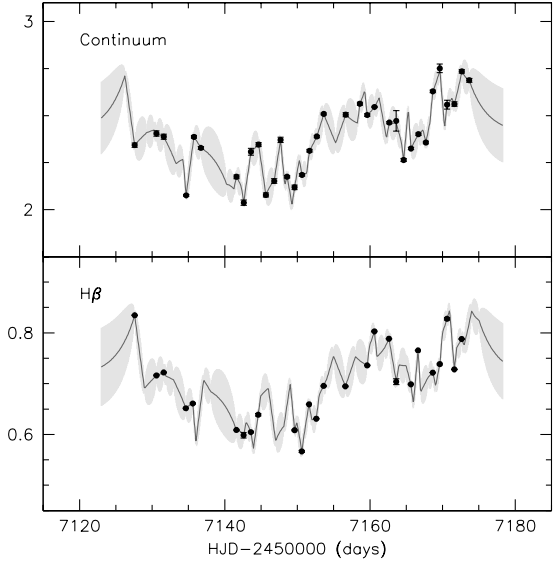}
\caption{Continuum and H$\beta$ light curves with interpolated JAVELIN
  light curves drawn from the distribution of acceptable models.}
\label{fig:jav}
\end{figure}

We also investigated the time delays with the {\tt JAVELIN} package
\citep{zu11}.  {\tt JAVELIN} fits the continuum variations with a
damped random walk model.  It then assumes a top hat model for the
reprocessing function, and determines the best-fit shifting and
smoothing parameters for the emission-line light curves by maximizing
the likelihood of the model.  Uncertainties on each of the model
parameters are assessed through a Bayesian Markov Chain Monte Carlo
method.  We denote time delays from {\tt JAVELIN} as $\tau_{\rm jav}$.
Given the extremely short time delays, we were unable to fit a single
model while including all the emission lines simultaneously, so we
instead modeled each emission line separately relative to the
continuum (see Figure~\ref{fig:jav}).

Time delay measurements are listed in Table~\ref{tab:lags}.  While
each of the measurements is an observed time delay, the rest-frame
time delays, corrected for a factor of $1+z$, are formally the same
within the uncertainties.

\begin{deluxetable}{lccc}
\tablecolumns{4}
\tablewidth{0pt}
\tablecaption{Time Lags}
\tablehead{
\colhead{Feature} &
\colhead{$\tau_{\rm cent}$} &
\colhead{$\tau_{\rm peak}$} &
\colhead{$\tau_{\rm jav}$} 
\\
\colhead{} &
\colhead{(days)} &
\colhead{(days)} &
\colhead{(days)}
}
\startdata
H$\beta$        & $1.4^{+0.7}_{-0.8}$  & $1.1^{+0.6}_{-0.6}$  & $1.3^{+0.2}_{-0.7}$  \\
H$\gamma$       & $0.0^{+1.0}_{-1.3}$  & $-0.7^{+2.5}_{-0.7}$ & $-1.5^{+0.1}_{-0.7}$ \\
\ion{He}{2}     & $-0.2^{+0.9}_{-1.1}$ & $-0.7^{+1.8}_{-0.7}$ & $-1.4^{+0.2}_{-0.1}$ 
\label{tab:lags}
\enddata 
\end{deluxetable}

\section{Line Width Measurements}

The widths of the broad emission lines in AGN spectra are interpreted
as the line-of-sight velocities of the bulk motion of the gas.  The
narrow emission lines, however, are known to emit from gas that is not
participating in the same bulk motion.  Therefore, good practice is to isolate
the broad emission from the narrow emission when quantifying
the line width.  In the spectrum of UGC\,06728, however, it is not
clear what part of the H$\beta$ line is narrow emission
(cf. Figure~\ref{fig:mean.rms}).  Furthermore, the narrow lines
contribute almost no signal to the rms spectrum, demonstrating that
our internal spectral scaling method has minimized their apparent
variability from changing observing conditions throughout the
monitoring campaign.  As it is the variable part of the emission line
(the rms profile) that we are most interested in, we do not attempt
any narrow line subtraction for this object.

We measured the widths of the broad H$\beta$, \ion{He}{2}\,$\lambda
4686$, and H$\gamma$ emission lines in both the mean and the rms
spectra and we report two different line width characterizations: the
full width at half the maximum flux (FWHM) and the second moment of
the line profile ($\sigma_{\rm line}$).  Line widths were measured
directly from the spectra, with each line profile defined as the flux
above a local linear continuum.  Uncertainties in the emission line
widths were determined using a Monte Carlo random subset sampling
method.  From a set of $N$ spectra, a subset of $N$ spectra were
selected without regard to whether they had been previously chosen.
The mean and rms of the subset were created, from which the FWHM and
$\sigma_{\rm line}$ of an emission line were determined and recorded.
Distributions of line width measurements were built up over 1000
realizations.  We take the mean and the standard deviation of each
distribution as the measurement and its uncertainty, respectively.

Following \citet{peterson04}, we corrected the emission-line widths
for the dispersion of the spectrograph.  The observed emission line
width, $\Delta \lambda_{\rm obs}$, can be described as
\begin{equation}
\Delta \lambda_{\rm obs}^2 \approx \Delta \lambda_{\rm true}^2 + \Delta \lambda_{\rm disp}^2
\end{equation}
\noindent where $\Delta \lambda_{\rm true}$ is the intrinsic line
width and $\Delta \lambda_{\rm disp}$ is the broadening induced by the
spectrograph.  The employment of a wide spectrograph slit for
reverberation campaigns means that the spectrograph dispersion cannot
be determined from night sky emission lines or from arc lamp lines ---
the unresolved AGN point source, even under poor seeing conditions,
will not fill the spectrograph slit.  Given the relative obscurity of
this particular AGN, we were unable to estimate $\Delta \lambda_{\rm
  true}$, and therefore constrain $\Delta \lambda_{\rm disp}$, from
high-quality, high-resolution observations of the narrow emission
lines in the literature.  However, we have previously monitored other
AGNs with this same instrumental setup, and so we adopt the value of
$\Delta \lambda_{\rm disp} = 14.1$\,\AA\ that we determined for
NGC\,5273 from a spring 2014 monitoring campaign \citep{bentz14}.

Our final resolution-corrected line width measurements are listed in
Table~\ref{tab:linewidths}.

\begin{deluxetable*}{lcccc}
\tablecolumns{5}
\tablewidth{0pt}
\tablecaption{Line Widths}
\tablehead{
\colhead{} &
\multicolumn{2}{c}{Mean} &
\multicolumn{2}{c}{RMS}
\\
\colhead{Feature} &
\colhead{FWHM} &
\colhead{$\sigma_{\rm line}$} &
\colhead{FWHM} &
\colhead{$\sigma_{\rm line}$} 
\\
\colhead{} &
\colhead{(km\,s$^{-1}$)} &
\colhead{(km\,s$^{-1}$)} &
\colhead{(km\,s$^{-1}$)} &
\colhead{(km\,s$^{-1}$)}
}
\startdata
H$\beta$     & $1144.5 \pm 58.3$  & $758.3 \pm 19.4$   & $1309.7 \pm 182.2$  & $783.7 \pm  92.3$  \\
H$\gamma$    & $2333.6 \pm 80.3$  & $821.8 \pm 21.8$   & $2492.3 \pm 1704.7$ & $919.9 \pm 70.4$   \\
\ion{He}{2}  & $2626.2 \pm 593.7$ & $1124.7 \pm 127.7$ & $4016.7 \pm 912.9$  & $1605.6 \pm 157.8$ 
\label{tab:linewidths}
\enddata 
\end{deluxetable*}

\section{Black Hole Mass}

All of the time delays measured for UGC\,06728 are very short, which
is to be expected given the low luminosity of the AGN.  The time
delays determined for H$\beta$ are the only ones that are not formally
consistent with zero within the measurement uncertainties, so H$\beta$
is the only emission line we will consider for the determination of
the black hole mass.  However, H$\beta$ is also the emission line for
which we have the largest number of reverberation results
(cf. \citealt{bentz15} for a recent summary), so it is also the most
reliable emission line for determining \mbh.

The black hole mass is generally determined from reverberation-mapping
measurements as:
\begin{equation}
M_{\rm BH} = f \frac{c \tau V^2}{G}
\end{equation}
where $\tau$ is the time delay for a specific emission line relative
to variations in the continuum, and $V$ is the line-of-sight velocity width
of the emission line, with $c$ and $G$ being the speed of light and
gravitational constants, respectively.  The emission-line time delay
is interpreted as a measure of the responsivity-weighted average radius of
the broad-line region for that specific emission feature (e.g.,
H$\beta$).

The scaling factor $f$ accounts for the detailed geometry and
kinematics of the broad line region gas, which is unresolvable.  In
practice, the multiplicative factor, $\langle f \rangle$, which is
found to bring the \msigma\ relationship for AGNs with reverberation
masses into agreement with the \msigma\ relationship for nearby
galaxies with dynamical black hole masses (e.g.,
\citealt{gultekin09,mcconnell13,kormendy13}) is used as a proxy for
$f$.  In this way, the population average factor provides an overall
scale for reverberation masses that should be unbiased, but the mass
of any particular AGN is expected to be uncertain by a factor of 2-3
because of object-to-object variations.  The value of $\langle f
\rangle$ has varied in the literature from 5.5 \citep{onken04} to 2.8
\citep{graham11}, depending on which objects are included and the
specifics of the measurements.  We adopt the value determined by
\citet{grier13} of $\langle f \rangle = 4.3 \pm 1.1$.

Combining the time lag ($\tau_{\rm cent}$) and line width
($\sigma_{\rm line}$) measurements for H$\beta$ and scaling by
$\langle f \rangle$, we determine $M_{\rm BH} = (7.1 \pm 4.0) \times
10^5$\,M$_{\odot}$.

\begin{figure}
\epsscale{1.1}
\plotone{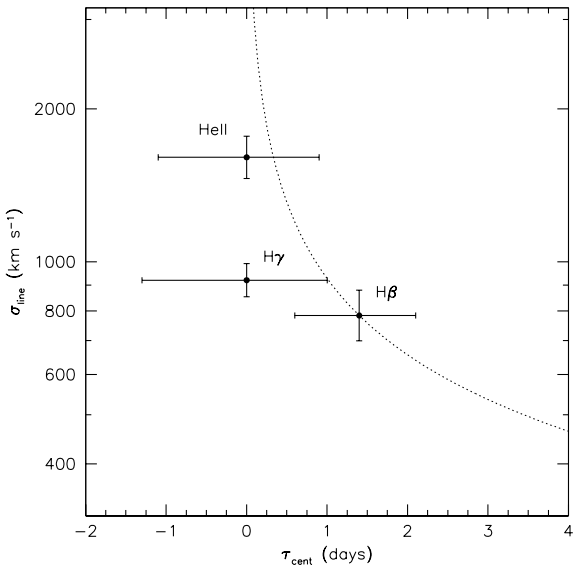}
\caption{Line width versus time delay as measured from the broad
  optical recombination lines in the spectrum of UGC\,06728.  The
  dotted line shows the expected relationship of $R \propto V^{-2}$
  and is scaled to match the measurements for H$\beta$.  Even though
  the time delays are quite short, and unresolved in the case of
  H$\gamma$ and \ion{He}{2}, the measurements are in relatively good
  agreement with the expected relationship.}
\label{fig:rv}
\end{figure}

\begin{figure*}
\epsscale{1.1}
\plotone{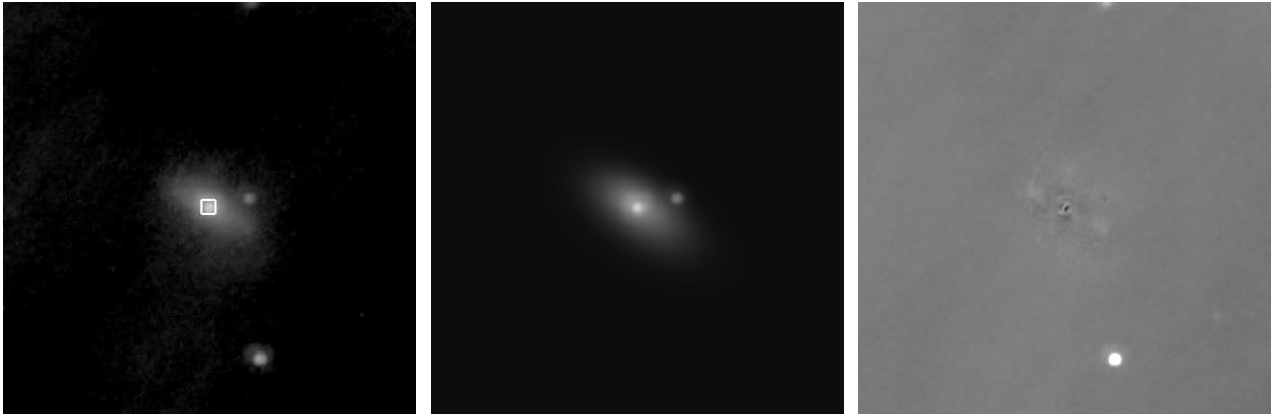}
\caption{Stacked $g-$band image of a 2\farcm5$\times$2\farcm5 region
  centered on UGC\,06728 ({\it left}) with the white rectangle showing
  the geometry of the ground-based spectroscopic monitoring aperture.
  The best-fit model determined from {\tt GALFIT} is displayed in the
  middle panel, and the right panel shows the residuals after
  subtraction of the model from the image. All images are oriented
  with north up and east to the right.}
\label{fig:galfit}
\end{figure*}

\section{Discussion}

The extremely rapid response of the broad emission lines to variations
in the continuum flux in UGC\,06728 means that our daily sampling was
not fine enough to resolve time delays for all the broad optical
recombination lines.  The time delay of H$\beta$ is the only one that
is not formally consistent with zero delay, and it is only marginally
resolved at that.  However, while we were not able to resolve the time
delays for H$\gamma$ and \ion{He}{2}, we can examine them in light of
the expected virial relationship for BLR gas that is under the
gravitational dominance of the black hole.  In particular, we would
expect that $R \propto V^{-2}$.  This relationship has been shown to
be a good description of observations when reverberation results from
multiple emission lines have been recovered (e.g.,
\citealt{peterson04,kollatschny03,bentz10a}).  Figure~\ref{fig:rv}
shows the measurements for the optical recombination lines in
UGC\,06728, with the expected relationship scaled to match the
measurements for H$\beta$.  There is generally good agreement with the
expected relationship within the measurement uncertainties, such that
we would not expect to resolve the responses of these emission lines
with our current sampling.  A monitoring campaign with finer temporal
resolution ($\Delta t = 0.25-0.5$\,days) would be needed to further
improve upon these constraints.

\subsection{Consistency with the \rl\ Relationship}

Furthermore, we can examine the location of UGC\,06728 on the AGN
\rl\ relationship to further assess the H$\beta$ time delay
measurement.  For very nearby galaxies like UGC\,06728, however, one
complication is the large fraction of host-galaxy starlight that
contributes to the continuum emission at rest-frame 5100\,\AA\ through
the large spectroscopic slit ($\sim 5$\arcsec) employed in a
reverberation mapping campaign.  The usual method to correct for this
contamination is to carry out two-dimensional surface brightness
modeling of a high-resolution image of the galaxy (usually from the
{\it Hubble Space Telescope} to maximize the image quality), thereby
isolating the host-galaxy starlight components from the unresolved AGN
point source.  Using the modeling results to create an ``AGN-free''
image allows the starlight contribution to be directly constrained
\citep{bentz06a,bentz09b,bentz13}.  Unfortunately, there are no {\it
  HST} images of UGC\,06728.  The highest resolution optical images
available are the APO DIS $g-$band images discussed above, with a
pixel scale of $0.42$\arcsec/pixel.  While hardly comparable to the
quality afforded by {\it HST}, the DIS images do allow us to place
some rough constraints on the starlight contribution to the flux
density at $5100\times(1+z)$\,\AA.

We aligned and stacked several of the $g-$band images to increase the
signal-to-noise in the combined image.  Using the two-dimensional
surface brightness fitting program {\tt GALFIT} \citep{peng02,peng10},
we created a model of the point spread function (PSF) of the stacked
image by fitting multiple Gaussian components to the profile of a
field star in a restricted portion of the image.  We then employed
this model PSF while fitting the full frame, including a background
sky gradient, a PSF for the AGN and the nearby star, an exponential
profile for the disk of the galaxy, and a \sersic\ profile for the
bulge.  The bulge profile, in particular, is very compact with a
half-light radius of 1.7\,pix (0.7\arcsec), and likely degenerate with
the AGN PSF, so we caution that our estimate of the starlight
contribution is probably more like a lower limit.
Figure~\ref{fig:galfit} displays a 2\farcm5$\times$2\farcm5 region of
the stacked $g-$band image, our best-fit model from {\tt GALFIT}, and
the residuals after subtracting the model from the image.

As described earlier, calibrated $g-$band photometry for three field
stars from APASS (the AAVSO Photometric All Sky Survey;
\citealt{henden14}) was used to set the overall flux scale of the
image.  We also account for a slight flux scaling factor, due to the
difference in effective wavelength of the $g$ filter compared to
$5100\times(1+z)$\,\AA, using {\tt Synphot} and a template galaxy
bulge spectrum \citep{kinney96}.  Our estimate of the host-galaxy
contribution to the spectroscopic flux density is $f_{\rm gal} = (1.09
\pm 0.22) \times 10^{-15}$\,ergs\,s$^{-1}$\,cm$^{-2}$\,\AA$^{-1}$.
Removing this contribution results in an AGN-only continuum flux
density of $f_{\rm AGN} = (1.12 \pm 0.23) \times
10^{-15}$\,ergs\,s$^{-1}$\,cm$^{-2}$\,\AA$^{-1}$.  Assuming a
luminosity distance of $D_L = 27$\,Mpc and correcting for Galactic
absorption along the line of sight \citep{schlafly11}, we derive $\log
\lambda L_{\lambda} = 41.83 \pm 0.24$\,ergs\,s$^{-1}$.

\begin{figure}
\epsscale{1.1}
\plotone{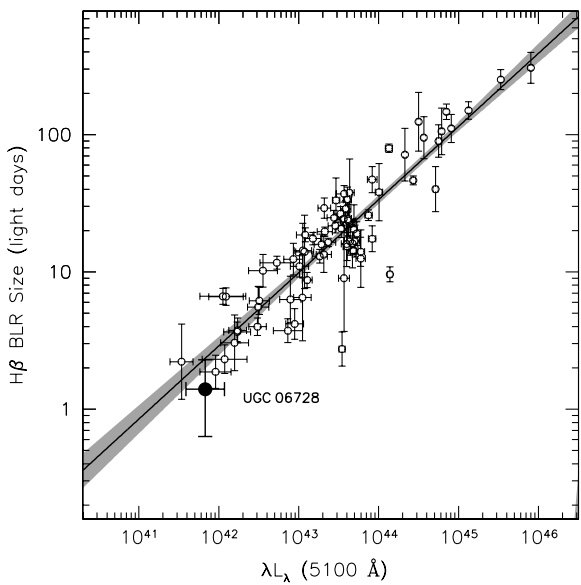}
\caption{The H$\beta$ time delay for UGC\,06728 and estimated AGN
  luminosity (filled point) compared to the radius-luminosity
  relationship for other reverberation-mapped AGNs \citep{bentz13}. }
\label{fig:rl}
\end{figure}

Figure~\ref{fig:rl} displays the \rl\ relationship for nearby AGNs
based on reverberation mapping of H$\beta$ \citep{bentz13}.  The
filled circle shows the location of UGC\,06728 with the H$\beta$ time
delay we have derived here and the luminosity after correction for the
estimated starlight contribution.  The agreement between UGC\,06728
and its expected location based on its estimated luminosity is
extremely good considering the barely-resolved nature of the time
delay and the caveats in the luminosity determination.  Furthermore,
we can expect that the agreement is actually somewhat better than
depicted, given the likelihood that the starlight correction to the
luminosity is underestimated as described above.

Taking our galaxy decomposition at face value, we can estimate the
bulge-to-total ratio as $B/T \approx 0.2$, which suggests that the
Hubble type of the galaxy is $\sim$Sb \citep{kent85}.  We also
estimate the color of the galaxy as $g-r \approx 0.9$, which suggests
$M/L_g \approx 6$ \citep{zibetti09}.  The total stellar mass of the
galaxy is $M_{\star} \approx 7.5\times10^9$\,M$_{\odot}$, which also
agrees with the host-galaxy being Sb$-$Sc in type.

\subsection{Consistency with the \msigma\ Relationship}

To further explore the reverberation results for UGC\,06728 within the
context of the larger reverberation sample, we obtained supplemental
observations on 2016 May 13 with the DIS Spectrograph on the APO 3.5-m
telescope with the intent of constraining the bulge stellar velocity
dispersion.  The high resolution B1200 and R1200 gratings were
employed, providing nominal dispersions of 0.62\,\AA/pix and
0.58\,\AA/pix and wavelength coverages of 1240\,\AA\ and 1160\,\AA,
respectively.  The blue grating was centered at 4900\,\AA\ to target
the Mg$b$ stellar absorption signature, and the red grating was
centered at 8500\,\AA\ for the \ion{Ca}{2} triplet absorption.  The
$0\farcs9$ slit was rotated to a position angle of 150\degr\ east of
north, approximately along the minor axis of the galaxy.  Given the
high inclination of the galaxy, we specifically avoided the major axis
of the galaxy to mitigate the effects of rotational broadening from
the disk within the one-dimensional extracted spectra.  Two 1200\,s
exposures were obtained through patchy clouds and with marginal seeing
at an airmass of 1.6.  Spectra of the standard star, Feige\,34, were
also obtained to assist with the flux calibration, as well as spectra
of HD\,125560 (spectral type K3III) and HD\,117876 (spectral type
G8III) to provide velocity templates with the same wavelength coverage
and dispersion as the galaxy.  All spectra were reduced with IRAF
following standard procedures.  An extraction width of 40\,pixels
(corresponding to 16\arcsec\ on the blue camera and 16.8\arcsec\ on
the red camera) was adopted to maximize the galaxy signal in the
resultant spectra.

\begin{figure}
\epsscale{1.1}
\plotone{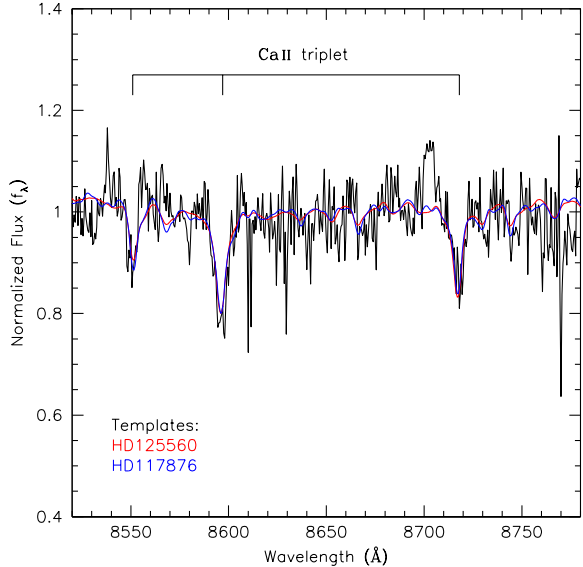}
\caption{Spectrum of UGC\,06728 in the wavelength region around the
  \ion{Ca}{2} triplet absorption lines.  The red and blue lines show
  the best-fit models to the stellar absorption lines based on
  HD125560 and HD117876, respectively.  We take the average of the
  solutions provided by the two template stars as our measurement of
  the bulge stellar velocity dispersion in UGC\,06728.  }
\label{fig:veldisp}
\end{figure}

Following flux calibration of the spectra, we employed the pPXF
(Penalized Pixel Fitting) method of \citet{cappellari04} to extract
the stellar kinematics.  The Mg$b$ absorption signature was not
detected in the galaxy spectra, but the \ion{Ca}{2} triplet features were
detected, so we focused on fitting the red spectra only.  During the
fitting process, we restricted the wavelength region to
8525$-$8850\,\AA\ and determined the best-fit parameters (velocity,
velocity dispersion, h3, and h4) using first one velocity template
star and then the other.  The best fits to the spectrum of UGC\,06728
are displayed in Figure~\ref{fig:veldisp}: HD125560 (red line)
provided a best-fit velocity dispersion of 56.5\,km\,s$^{-1}$, and
HD117876 (blue line) provided a best fit of 46.7\,km\,s$^{-1}$.  We
take the average of these as the bulge stellar velocity dispersion,
$\sigma_{\star} = 51.6 \pm 4.9$\,km\,s$^{-1}$.

\begin{figure}
\epsscale{1.1}
\plotone{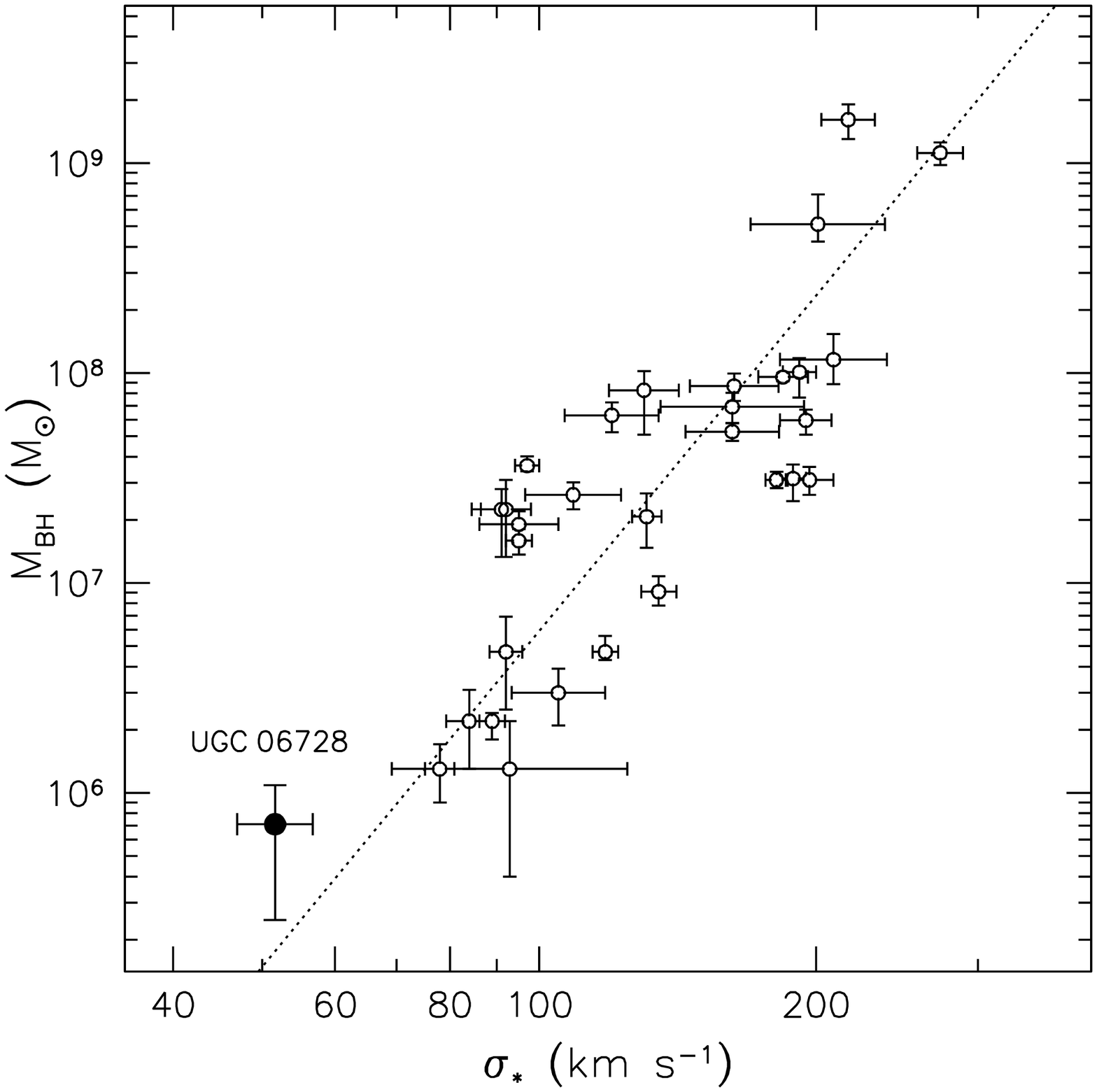}
\caption{UGC\,06728 (filled point) and the AGN \msigma\ relationship
  from \citet{grier13}.}
\label{fig:msig}
\end{figure}

With this constraint on the bulge stellar velocity dispersion in
UGC\,06728, we can explore its location on the AGN
\msigma\ relationship. Figure~\ref{fig:msig} displays the AGN
\msigma\ relationship from \citet{grier13} (open points and line),
with the location of UGC\,06728 shown by the filled circle.  The
scatter at the low-mass end of the \msigma\ relationship for AGNs with
reverberation masses seems to be much smaller than that found for
megamaser host galaxies \citep{greene10}.  \citet{lasker16} also found
the megamaser host galaxies to have a high scatter relative to the
\ml\ and \mm\ relationships.  Each sample of direct black hole masses,
whether dynamical, reverberation, or masering, has its own set of
biases and assumptions that are independent of the other techniques,
so further exploration into this apparent disagreement is likely to
shed light on the reliability of black hole mass measurements as they
are currently applied.

Furthermore, we can estimate the black hole sphere of influence
($r_h$) in the nucleus of UGC\,06728. Generally defined as
\begin{equation}
r_h = \frac{GM_{\rm BH}}{\sigma_{\star}^2}
\ ,
\end{equation}
\noindent $r_h$ is often employed as a convenient metric for
determining the probability of success for constraining $M_{\rm BH}$
from spatially resolved stellar dynamics.  \citet{gultekin09} argue
that a strict reliance on resolving $r_h$ is not necessary, however,
for useful constraints on black hole masses.

Combining our measurements of $M_{\rm BH}$ and $\sigma_{\star}$ and
again assuming a luminosity distance of $D_L=27$\,Mpc, we estimate $r_h
= 0.01$\arcsec\ for UGC\,06728.  While this angular size is smaller
than the achievable spatial resolution of integral field spectrographs
on the largest ground-based telescopes today, it is interesting to
note that it is not much smaller than $r_h$ for NGC\,3227.
\citet{davies06} were able to constrain the black hole mass of
NGC\,3227 through stellar dynamical modeling, even though the
reverberation mass and bulge stellar velocity dispersion predict $r_h
= 0.018$\arcsec.  Given the very limited number of AGNs where it will
be possible to carry out a direct comparison of reverberation-based
and stellar dynamical-based black hole mass measurements with current
and near-future technology, UGC\,06728 could potentially be a
worthwhile target for dynamical modeling.

\subsection{Mass and Spin Implications}

\citet{walton13} analyzed {\it Suzaku} observations of UGC\,06728 and
determined that it was a ``bare'' AGN, with minimal intrinsic
absorption.  Fitting the X-ray spectrum with a relativistic reflection
model, and assuming an accretion disk inclination of $i=45$\degr, they
determined a dimensionless spin parameter of $a > 0.7$, indicating the
black hole is spinning rapidly.  Combined with our mass contraint of
$M_{\rm BH} = (7.1 \pm 4.0) \times 10^5$\,M$_{\odot}$, UGC\,06728 is
one of a small number of massive black holes that are completely
characterized.  A few other low-mass black holes have both mass and
spin constraints, and they appear to agree with the properties derived
for UGC\,06728.  MCG-06-30-15 is only slightly more
massive with $M_{\rm BH}=(1.6\pm0.4)\times10^6$\,M$_{\odot}$
\citep{bentz16a} and is spinning near maximally ($a > 0.9$;
\citealt{brenneman06,chiang11,marinucci14}).  NGC\,4051 is another
example, with $M_{\rm BH}=(1.3\pm0.4)\times10^6$\,M$_{\odot}$
\citep{denney09b} and $a>0.99$ \citep{patrick12}.

Black hole evolutionary models have only recently begun to treat black
hole spin in addition to mass.  Depending on the model, it is not
clear if the properties of the black hole in UGC\,06728 are expected
or surprising.  For example, the model of \citet{volonteri13} predicts
that black holes with $M_{\rm BH} \approx 10^6$\,M$_{\odot}$ in
gas-rich galaxies at $z < 0.5$ (including AGNs) should have slowly
rotating black holes with dimensionless spin parameters of $a < 0.4$.
This model is based on many observational constraints, including the
\msigma\ relationship, with which we have shown UGC\,06728 to be in
agreement.  One caveat to the evolutionary model of
\citet{volonteri13} is that it does not account for black hole feeding
through disk instabilities, which could be a reason for the apparent
discrepancy here. Disk instability accretion events would likely be
correlated and serve to spin up a black hole.  The evolutionary models
of \citet{sesana14} attempt to include this effect by linking the gas
dynamics of the extended galaxy to the central black hole.  Their
models predict that local black holes with $M_{\rm BH} \approx
10^6$\,M$_{\odot}$ should tend to be spinning near maximally, and that
{\it accreting} black holes in spiral galaxies should also tend to
have near-maximal spins.

Interpretation of black hole spin measurements is still somewhat
debated as well. \citet{bonson16} argue that black hole spins tend to
be overestimated in many cases, although they state this is likely not
the case for the most maximally spinning black holes ($a > 0.8$).
Furthermore, there is a very strong selection bias inherent in the
sample of AGNs with spin measurements.  Rapidly spinning black holes
have significant boosts to their X-ray flux through increased
radiative efficiency, and the current sample of AGNs with spin
constraints is based on observations of the brightest X-ray sources,
so the current sample will strongly favor rapidly spinning black holes
\citep{brenneman11,vasudevan16}.  In any case, UGC\,06728 is an
important addition to the sample. As the least massive central
black hole that has been fully described, it will help to anchor
future studies, both observational and theoretical, of central black
hole {\mbox demographics.}

\section{Summary}

We present an H$\beta$ time delay and a reveberation-based black hole
mass for the nearby, low-luminosity Seyfert UGC\,06728.  With
$\tau=1.4\pm0.8$\,days and $M_{\rm BH} = (7.1 \pm 4.0) \times
10^5$\,M$_{\odot}$, UGC\,06728 is at the low end of observed
properties within the reverberation mapping sample.  The time delay
and estimated AGN luminosity agree with the \rl\ relationship for
other reveberation-mapped AGNs, and a measurement of $\sigma_{\star} =
51.6 \pm 4.9$\,km\,s$^{-1}$ from long-slit spectroscopy shows that the
black hole mass agrees with the AGN \msigma\ relationship.  With
$M_{\rm BH}<10^6$\,M$_{\odot}$, UGC\,06728 is currently the
lowest-mass central black hole that is fully described by both direct
mass and spin constraints.

\acknowledgements

We thank the referee for thoughtful comments that improved the
presentation of this paper.  MCB gratefully acknowledges support from
the NSF through CAREER grant AST-1253702.  This research is based on
observations obtained with the Apache Point Observatory 3.5-meter
telescope, which is owned and operated by the Astrophysical Research
Consortium.  We heartily thank the staff at APO for all their help
with this program.  This research has made use of the AAVSO
Photometric All-Sky Survey (APASS), funded by the Robert Martin Ayers
Sciences Fund.  This research has made use of the NASA/IPAC
Extragalactic Database (NED) which is operated by the Jet Propulsion
Laboratory, California Institute of Technology, under contract with
the National Aeronautics and Space Administration and the SIMBAD
database, operated at CDS, Strasbourg, France.


\end{document}